\documentclass[12pt]{iopart}

\expandafter\let\csname equation*\endcsname\relax
\expandafter\let\csname endequation*\endcsname\relax
\usepackage{amsmath} %mathematical environments and symbols
\usepackage{amsfonts}
\usepackage{amssymb}
\usepackage{siunitx} %mainly for producing the degrees symbol using \ang{}
\usepackage{braket} %for BraKet notation
\usepackage{graphicx} %general figures stuff
\usepackage{nicefrac}
\usepackage{hyperref} %in-document links
\usepackage{wrapfig} %figure placement
\usepackage{caption}
\usepackage{subcaption} %subcaptions for subfigures
\usepackage{fancyhdr} %for fancy headers
\usepackage[utf8]{inputenc} %related to encoding
\usepackage{etoolbox} %removes extra 'references title when calling bib group
\patchcmd{\thebibliography}{\section*{\refname}}{}{}{}%same as above
\usepackage{float} %fix figure positions with [H] (stronger than  [h!])
\usepackage{parskip} % OPTIONAL - removes paragraph indentation
\graphicspath{{img/}{../img/}} %sets path for images/figures
\usepackage{cite}
\begin{document}
%\NJP

\title{Localisation determines the optimal noise rate for quantum transport}
\author{Alexandre R. Coates$^1$, Brendon W. Lovett$^2$ and Erik M. Gauger$^1$}\address{$^1$ SUPA, Institute of Photonics and Quantum Sciences, Heriot-Watt University, Edinburgh EH14 4AS, United Kingdom} \address{$^2$ SUPA, School of Physics and Astronomy, University of St Andrews, St Andrews KY16 9SS, United Kingdom} 
\ead{ac173@hw.ac.uk}

\date{\today}
\begin{abstract}
Environmental noise plays a key role in determining the efficiency of transport in quantum systems. However, disorder and localisation alter the impact of such noise on energy transport. To provide a deeper understanding of this relationship we perform a systematic study of the connection between eigenstate localisation and the optimal dephasing rate in 1D chains. The effects of energy gradients and disorder on chains of various lengths are evaluated and we demonstrate how optimal transport efficiency is determined by both size-independent, as well as size-dependent factors. By discussing how size-dependent influences emerge from finite size effects we establish when these effects are suppressed, and show that a simple power law captures the interplay between size-dependent and size-independent responses. Moving beyond phenomenological pure dephasing, we implement a finite temperature Bloch-Redfield model that captures detailed balance.
We show that the relationship between localisation and optimal environmental coupling strength continues to apply at intermediate and high temperature but breaks down in the low temperature limit.\\
\noindent{\it Keywords:\/ Quantum Transport, Disorder, Localisation, Open Quantum Systems,  Noise-Assisted Quantum Transport}
\end{abstract}

\maketitle
%\twocolumngrid
\section{Introduction}
\label{sec: intro}
Energy transport occurs in many contexts: from circuits and molecular junctions to processes like photosynthesis~\cite{Kassal2013DoesPhotosynthesis, Ishizaki2009UnifiedApproach, Engel2007EvidenceSystems, Brixner2017ExcitonSystems} and the electron transport chain in biology~\cite{Kundu2017NanoscaleHarvesting}. This fundamental process has very different features depending on the scale on which it acts and the specifics of the system coupling to the environment~\cite{Amarnath2016MultiscalePlants, Bennett2013ADescription}. For over a decade, a lot of work has exposed the mechanisms of Environmental Noise-Assisted Quantum Transport (ENAQT)~\cite{Plenio2008Dephasing-assistedBiomolecules, Mohseni2008Environment-AssistedTransfer, Chin2010Noise-assistedComplexes, Zerah-Harush2018UniversalNetworks, Dwiputra2021Environment-assistedEdges}, a phenomenon describing how incoherent processes from interactions with the environment around a system can improve energy transport in quantum systems. This work was heavily motivated by the possible connection between ENAQT and the efficiency of photosynthesis~\cite{Kassal2013DoesPhotosynthesis,Engel2007EvidenceSystems, Plenio2008Dephasing-assistedBiomolecules,Mohseni2008Environment-AssistedTransfer, Chin2010Noise-assistedComplexes,Lambert2012QuantumBiology, Huelga2013VibrationsBiology, Stones2016VibronicTransfer}, though recent work suggests the relationship between the two may be more nuanced~\cite{Harush2021DoNot, Higgins2021PhotosynthesisTransfer, Duan2017NatureTransfer.}.

\begin{figure}[H]
    \centering
    \includegraphics[width = .95\linewidth]{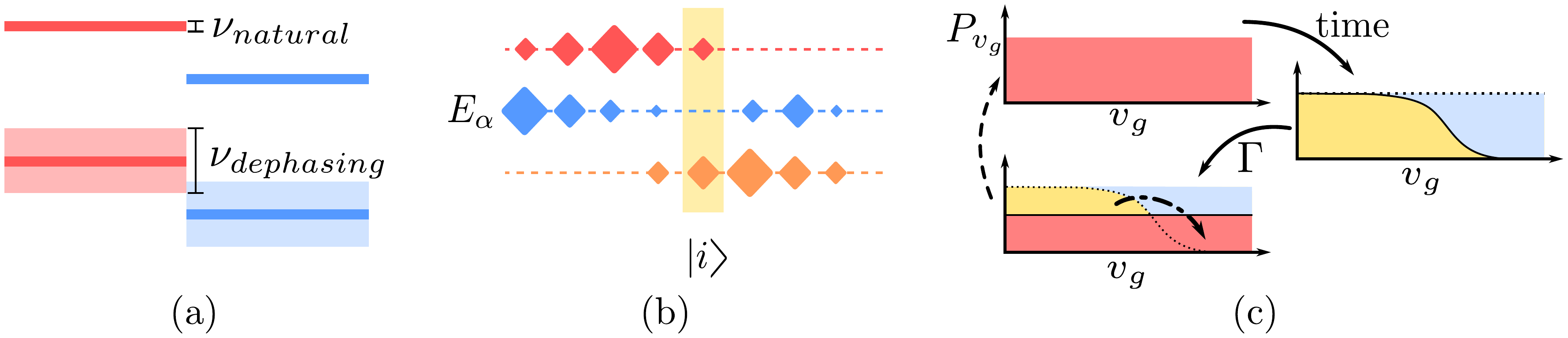}
    \caption{Illustrations of the ENAQT mechanisms that are relevant in this paper: dephasing induced line broadening (a), the invariant subspace (b) and momentum rejuvenation (c). Dephasing (and other forms of decoherence) act to broaden the linewidth of system states, making otherwise forbidden transitions energetically possible, which enables faster energy transport in disordered systems. The invariant subspace describes the eigenstates of a coupled system that have zero overlap with the particular energy extraction site $\ket{i}$; disorder and localisation increases the extent of this subspace, and environmental noise is needed to access it from the extraction site. Momentum rejuvenation is a finite size effect: it describes how high group velocity components of a population leave a system first, producing a skewed velocity distribution. Incoherent noise resets the distribution, effectively pumping population from low to high group velocities.}
    \label{fig: ENAQT schematic}
\end{figure}

There are a number of different ways in which ENAQT can arise, as shown in figure \ref{fig: ENAQT schematic}. These include line broadening which can help to overcome energetic barriers; the breaking up of an `invariant subspace' of the system Hamiltonian that is inaccessible to extraction operators on a quantum system~\cite{Chin2010Noise-assistedComplexes}; and momentum rejuvenation which counteracts the tendency of a fraction of the excitation to get stuck in only sluggishly propagating states~\cite{Li2015MomentumFlow}. Recent studies of steady state populations have also shown that the occupation of system sites becomes more uniform when transport efficiency is near-optimal~\cite{Zerah-Harush2018UniversalNetworks,  Dwiputra2021Environment-assistedEdges,Zerah-Harush2020EffectsTransport}, this population uniformisation phenomenon is discussed in \ref{sec: uniformisation}.

\begin{figure}[H]
    \centering
    \includegraphics[width = .95\linewidth]{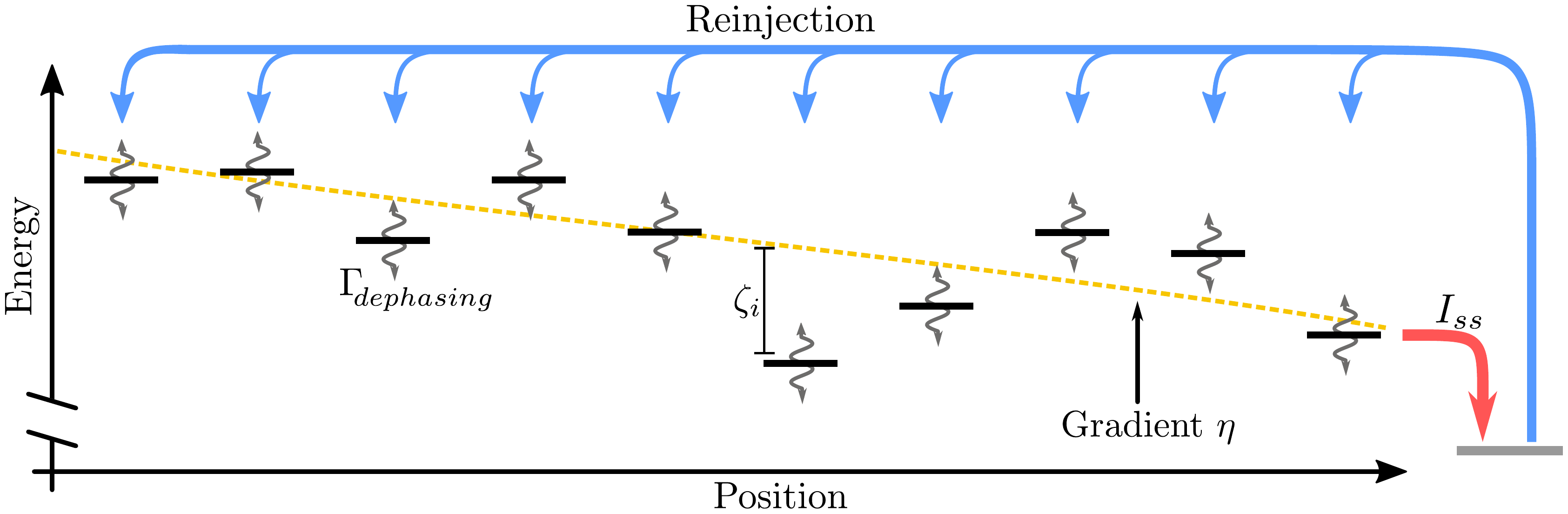}
    \caption{Schematic view of our system setup, showing a chain with ten sites of different energies, the energy of each site is altered by some random disorder $\zeta_i$ and a uniform and linear energy gradient $\eta$. Coupling to an environment induces dephasing $\Gamma$ on each site. The measure of transport efficiency that we use is the steady state current $I_{ss}$ extracted from the last site. After extraction the chain population is reinjected back onto all sites equally. Our goal is to find $\Gamma_{optimal}$ where $I_{ss}$ is maximised for the given combination of $\eta$ and $\zeta$.}
    \label{fig: chain schematic}
\end{figure}

In this paper, we perform a systematic study of how localising the eigenstates of 1D chains modifies their transport efficiency. Figure \ref{fig: chain schematic} illustrates the model we will consider here, which allows us to study three mechanisms that limit the delocalisation of chain eigenstates: limiting the total length of the chain, introducing static disorder, and applying a uniform energy gradients. Varying static disorder induces Anderson localisation~\cite{Anderson1958AbsenceLattices}, while a linear energy gradient produces Wannier-Stark localisation~\cite{Wannier1960WaveField}. 

%The effects of disorder on ENAQT have been studied before for a few ENAQT mechanisms. 

Previous studies on the effects of disorder on ENAQT have focused on how disorder affects the extent of the invariant subspace~\cite{Chin2010Noise-assistedComplexes, Caruso2009HighlyTransport} as well as the distribution of steady-state populations~\cite{Zerah-Harush2020EffectsTransport}. These studies have consistently found that as static disorder increases, more dynamic disorder is needed to improve transport efficiency\cite{Chin2010Noise-assistedComplexes, Zerah-Harush2020EffectsTransport}. More static disorder means more pure dephasing is needed to enable otherwise forbidden transitions, therefore the optimal pure dephasing rate is generally positively correlated with static disorder. 

Momentum rejuvenation, unlike other ENAQT mechanisms is a finite size effect~\cite{Li2015MomentumFlow}. High group velocity components of a propagating wave-packet explore and quickly exit the finite sized system, leaving behind a skewed velocity distribution which can be reset by environmental noise, repopulating the depleted higher velocity states. A consequence of this mechanism is that larger systems need longer before faster exciton components can escape, therefore they need to be `reset' less often, meaning the optimal noise rate is reduced.

In this paper we aim to produce a deeper understanding of the relationship between ENAQT and localisation, and we will also show that momentum rejuvenation continues to apply in non-degenerate systems and in the steady state. This allows us to compare the effect disorder has on size-dependent and size-independent ENAQT mechanisms. 

The focus of this work is on chains with short-range nearest-neighbour coupling, as this model is widely studied and can be fully localised. Long-range coupling has been observed in relevant experimental systems such as molecular aggregates~\cite{Spano1991CooperativeAggregates, Gulli2019MacroscopicNanotubes, Strumpfer2012HowLight-harvesting} or ion traps~\cite{Jurcevic2014QuasiparticleSystem}. However, in general the long-range interactions in 1D systems prevent full Anderson localisation~\cite{Levitov1989AbsenceInteraction,Evers2008AndersonTransitions}, and recent work has shown that homogeneous long-range coupling~\cite{Celardo2016ShieldingHopping} or coupling to cavities~\cite{Chavez2020Disorder-EnhancedCavities} can significantly alter 1D responses to disorder in ways beyond the scope of this paper. Recent years have also seen broad interest in the transient effects of dephasing on quantum diffusion, such as stochastic resonance, and many-body localisation, especially focused on the quasiperiodic Aubry-André model~\cite{Gholami2017NoiseModels, Lorenzo2018RemnantsNoise, Zhu2021ProbingComputer, Malla2018SpinfulCoupling, Bonca2018DynamicsBaths,Dwiputra2021Environment-assistedEdges, Prelovsek2018TransientBosons}, as well as quantum chaotic systems~\cite{Deutsch2018EigenstateHypothesis, DAlessio2016FromThermodynamics, Sa2020ComplexChaos, Rubio-Garcia2021FromLiouvillians}. We find no non-trivial transient effects in our model (see \ref{sec: transient effects}), so there remains open question of how the findings presented here would apply to more complicated scenarios.
\section{Theoretical Model}
\label{sec: Theory}
\subsection{System Model}
\label{sec: model}

In this paper we model chains with the single excitation approximation, defining the Hamiltonian as 
\begin{equation}
    H = \sum_i \epsilon_i \ket{i}\bra{i} + J \sum_{i=1}^{N-1} \ket{i}\bra{i+1} + \text{H.c.},
    \label{eqn: Ham}
\end{equation}
$\ket{i}$ represents a state with single excitation, on site $i$. 
$\epsilon_i$ is the on-site energy for site $i$, H.c is the Hermitian conjugate, and $J$ is the strength of the coupling between neighbouring sites. For this work $\hbar = 1$ and all quantities are given in terms of the coupling strength $J$ so we can focus on capturing the influence of disorder and gradients in a very general sense.

We consider chains of $N$ sites with site energies $\epsilon_i$ determined by a combination of energy disorder and a gradient in average site energies. As a convention we set the $\epsilon_0$ and $\epsilon_N$ to the highest average energy and zero respectively, from this we can define $\eta = \frac{\epsilon_0 - \epsilon_N}{N \cdot J}$ which we use to define the effective gradient applied to our system, scaled by system length and given in terms of the coupling strength $J$. To each site energy we add a perturbation $\zeta(\sigma)_i$ drawn from a Gaussian distribution, centred on zero with a standard deviation $\sigma$; here, $\sigma$ denotes the disorder strength for the system. 

The three parameters of system size $N$, disorder strength $\sigma$ and gradient $\eta$ all help define eigenstates and their localisations. The disorder introduces energy gaps, constraining eigenstates through localisation~\cite{Anderson1958AbsenceLattices}. The gradient could be a result of the application of a field to the system, and produces Wannier-Stark localisation~\cite{Wannier1960WaveField, Kleinman1990CommentLocalization, Emin1987Phonon-assistedField, Wilkinson1996ObservationPotential}. To measure localisation we use the inverse participation ratio (IPR), which is a measure of the number of sites over which each eigenstate $E_\alpha$ is delocalised. The average IPR over all eigenstates is defined as
\begin{equation}
   \text{IPR} = \frac{1}{\sum_{i,\alpha} |\braket{i| E_{\alpha}}|^4}. 
   \label{eqn: IPR}
\end{equation}

\begin{figure}[H]
    \centering
    \includegraphics[width = .95\linewidth]{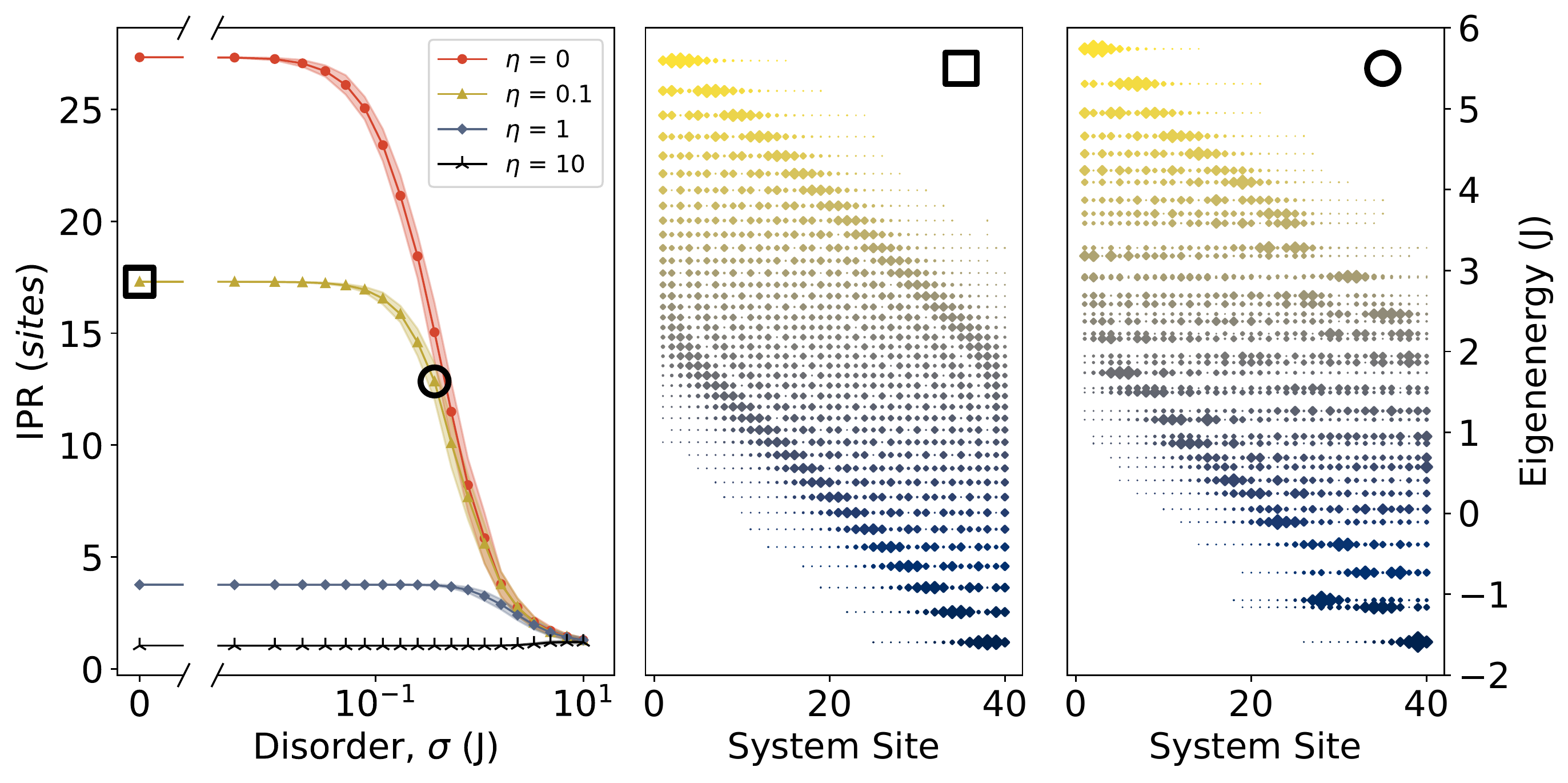}
    \caption{(Left) Average IPR against various disorder strengths $\sigma$ for $N = 40$, considered against four energy gradients $\eta$. Coloured areas show $\pm$ one standard deviation, each point is averaged from 100 configurations of disorder. An ordered ($\square$) and disordered ($\bigcirc$) point are highlighted, and their eigenspectra shown in the centre and right panels, respectively. (Centre) The eigenspectrum for a chain with $\eta = 0.1, \sigma = 0J$, showing the slight localisation of eigenstates under a uniform field. The size of each diamond is proportional to the probability of observing each eigenstate on that site. (Right) The eigenspectrum for $\eta = 0.1, \sigma = 0.3695J$, showing a mixture of field effects and disorder, producing inconsistent eigenenergy spacing as well as slightly more localised eigenstates.}
    \label{fig: ipr bias vs Anderson}
\end{figure}

This single value represents how localised that system is, with greater localisation implying not only a larger invariant subspace, but also a decreased efficiency in coherent transport. The IPR captures the system-wide impact of different gradients and disorder strengths on a system, making it a natural measure to compare systems. The effects of gradients and random disorder are illustrated in figure \ref{fig: ipr bias vs Anderson}, the coloured areas in the left panel show one standard deviation around the mean value at each point.
\subsection{Dynamics, Lindblad and Redfield Master Equations}
\label{sec: dynamics and rates}

We model each chain with a Lindblad master equation implemented with the QuTiP package~\cite{Johansson2013QuTiPSystems},

\begin{equation}
    \Dot{\rho} = -i [H,\rho] + 
    \Gamma \sum_{i = 1}^N \mathcal{L}\left[A_{deph, i}\right]\rho + \gamma_{inj}\sum_{i=1}^N\mathcal{L}\left[A_{inj, i}\right] +   \gamma_{trap}\mathcal{L}\left[A_{ext}\right]\rho, 
    \label{eqn: lindblad ME}
\end{equation}
where \(\mathcal{L}\left[A\right]\rho \) is the Lindbladian dissipator

\begin{equation}
    \mathcal{L}\left[A\right]\rho = \left(A \rho A^\dagger - \frac{1}{2} \{ A^\dagger A, \rho \} \right).
    \label{eqn: lindblad dephasing}
\end{equation}

\(\Gamma\) sets the rate of (dephasing) noise in the system, for simplicity assumed to be the same on each site, $\{\cdot,\cdot\}$ is the anticommutator, and $A_{deph,i}$ are Lindblad operators describing the environmental influence on each site $i$. For on-site dephasing in the single excitation approximation, the operators for on-site energy noise take the form $A_{deph,i} = 2\ket{i}\bra{i} - \mathbb{I}$~\cite{Jeske2015Bloch-RedfieldComplexes, Huo2012InfluenceSystems} . The extraction operator projects population from the $N^{\text{th}}$ site (lowest end of chain) to an external shelf state where it is trapped, $A_{ext} = { \sigma_N^+\sigma_{trap}^-}$. Similarly population is re-injected from the trap back onto each site with the injection operators $A_{inj, i} = { \sigma_{trap}^+\sigma_{i}^-}$. 

To treat these systems at finite temperatures we use the Bloch-Redfield master equation. As we study disordered systems with very mixed energy splittings we retain all non-secular terms to ensure it remains accurate~\cite{Eastham2016Bath-inducedApproximation}. The master equation reads:
\begin{equation}
    \begin{split}
        \Dot{\rho_s} &= -i [H,\rho_s] + \gamma_{inj} \sum_{i=1}^N\mathcal{L}\left[A_{inj, i}\right] +   \gamma_{trap} \mathcal{L}\left[A_{ext}\right]\rho \\
        &+ \Gamma \sum_{\omega} \sum_{m, n} S_{m,n}(\omega) \left( A_n(\omega)\rho_s A^\dagger_m(\omega) 
    - \frac{1}{2} \{ A^\dagger_m(\omega)A_n(\omega), \rho_s \} \right),
    \label{eqn: redfield ME}
    \end{split}
\end{equation}
where the injection and extraction operators are the same as in equation \ref{eqn: lindblad ME}, $\rho_s$ is the system density matrix and the frequencies $\omega$ are the eigenenergy splittings~\cite{Breuer2002TheSystems}. $A_{m}$ are system-environment interactions, derived by transforming the relevant site basis operators $A_{deph,i} = 2\ket{i}\bra{i} - \mathbb{I}$ into the Hamiltonian eigenbasis~\cite{Breuer2002TheSystems} and $S_{m n}(\omega)$ defines the noise-power spectrum associated with the system-environment interaction. The noise-power spectrum function is
\begin{equation}
    S_{m n}(\omega) =\left(\mathcal{N}_{BE}(\omega, \beta) + \Theta(\omega)\right)\mathcal{J}(\omega),
\end{equation}
where $\mathcal{N}_{BE}(\omega)$ defines Bose-Einstein statistics at a given phonon inverse temperature $\beta, \Theta(\omega)$ is the Heaviside function, allowing phonon-assisted transitions from higher to lower eigenenergies ($\omega > 0$) but not the reverse case, and $\mathcal{J}(\omega)$ is the spectral density. 
We use a flat spectral density as assumed in equation \ref{eqn: lindblad dephasing}, such that $\mathcal{J}(\omega) = \mathcal{J}$, this is for a direct comparison with the pure dephasing case. A Drude-Lorentz spectral density is considered and presented in \ref{sec: drude-lorentz-spectra}.

\subsection{Steady state setup and observables}
\label{sec: initials and observables}

As indicated by figure \ref{fig: chain schematic}, we re-inject any extracted population back onto all chain sites equally. By linearity, each injection site represents an initially populated site in the dynamical approach, so this injection scheme is equivalent to a mixed initial state. This choice ensures we capture the general system response, minimising the influence from inversion symmetry effects, while also ensuring that we can generically compare transport properties across systems with different sizes without adding in extra concerns about differing lengths between injection and extraction. For completeness, we show in \ref{sec: single-site injection} that injection on a single site produces qualitatively similar results.

We match the total injection to the extraction rate so that $\gamma_{inj} = \frac{\gamma_{trap}}{N}$. Our focus on steady-state properties is motivated by prior ENAQT studies which have shown that the steady state approach is more natural for energy transport in photosynthetic systems~\cite{Brumer2018SheddingProcesses, Axelrod2018AnLight, Kassal2013DoesPhotosynthesis}. 

The steady state $\rho_{ss}$ is found by calculating the zero eigenstate of the system Liouvillian. In our work $\gamma_{trap} = 3J$; changing this value generally changes quantitative values but not the qualitative behaviour~\cite{Zerah-Harush2020EffectsTransport} unless the rate is so high it begins to enter the Zeno regime~\cite{Chaudhry2016AEffects}. The key observable of transport efficiency is the steady state current $I_{ss}$, which we aim to maximise. This is simply the product of the extraction rate and the excited steady state population on the extraction site $N$,
\begin{equation}
    I_{ss} = \rho_{N,N} \gamma_{trap}.
    \label{eqn: steady state current}
\end{equation}

\section{Results}
\label{sec: results}
In this section we show how random disorder, energy gradients and system size affect ENAQT in the pure dephasing limit (section \ref{sec: dephasing}), demonstrating the strikingly consistent relationship between IPR and $\Gamma_{optimal}$. We also present a power law that fits the unbiased chain data, letting us separate the influence of size-independent and size-dependent effects. We then go beyond pure dephasing with the Bloch-Redfield master equation and show these effects are still qualitatively robust at high to intermediate temperatures, but break down in the lower temperature limit (section \ref{sec: bloch-redfield}).

\subsection{Pure Dephasing}
\label{sec: dephasing}

\begin{figure}[H]
    \centering
    \includegraphics[width = .8\linewidth]{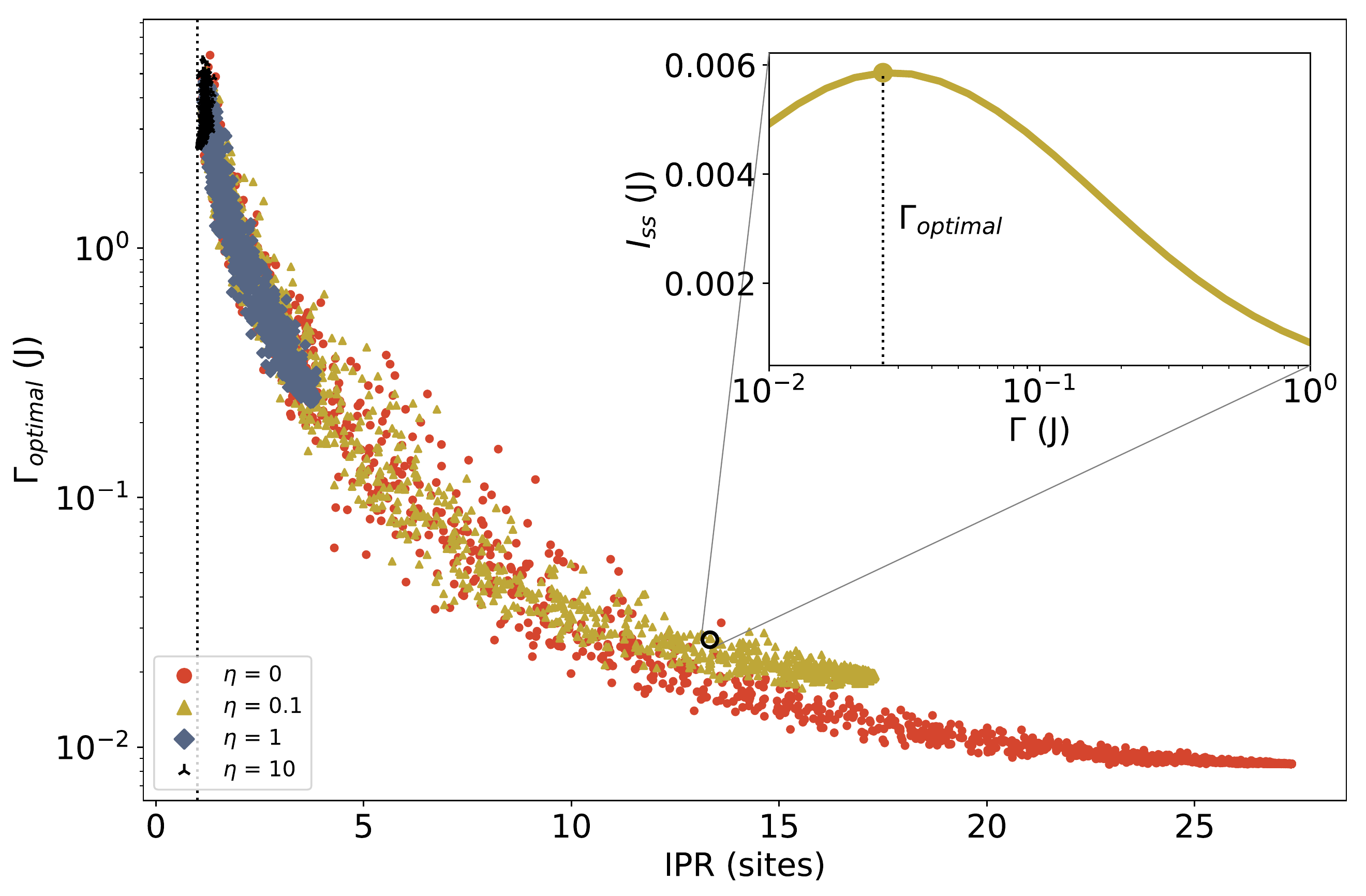}
    \caption{$\Gamma_{optimal}$ vs IPR for a variety of disordered $N$ = 40 chains, colour coded for each of the four different gradients $\eta$ considered. The inset shows how the curve points are generated: by varying the dephasing rate $\Gamma$ until a peak current is found. We see that in the majority of cases the trends overlap for each gradient, suggesting the IPR matters more than specific energy landscape or gradient. Calculations are repeated 100 times for each combination of gradient and disorder strength. The inset curve is calculated for $\eta = 0.1, \sigma = 0.3695J$ as in figure \ref{fig: ipr bias vs Anderson}.}
    \label{fig: N40 gradient comparison}
\end{figure}

Figure \ref{fig: N40 gradient comparison} shows how energy gradients and random disorder affect the optimal dephasing for chains of 40 sites. In all cases we see that as the IPR decreases linearly $\Gamma_{optimal}$ increases rapidly, once again confirming positive correlation between ENAQT peak position and static disorder~\cite{Chin2010Noise-assistedComplexes, Caruso2009HighlyTransport, Zerah-Harush2020EffectsTransport}. The main finding is that the results for each gradient largely overlap once the chain is disordered enough, indicating that once the system eigenstates are sufficiently localised the source of localisation does not matter.

Figure \ref{fig: N40 gradient comparison} shows that for sufficiently large IPRs (IPR $\geq 12$), the optimal dephasing for no gradient ($\eta = 0$) is lower than that for a weak gradient ($\eta = 0.1$). With momentum rejuvenation we expect that the larger the system is, the lower its $\Gamma_{optimal}$. As such, we infer that the presence of nonzero gradients limits the maximum length momentum rejuvenation can work over. So for $N = 40$ the gradient $\eta = 0.1$ is enough to slightly reduce the impact of momentum rejuvenation as compared to when $\eta = 0$. The result is a higher $\Gamma_{optimal}$ for the weak gradient.

As discussed in Sec. \ref{sec: Theory}, linear energy gradients localise eigenstates~\cite{Wannier1960WaveField,Emin1987Phonon-assistedField} and alter charge transport~\cite{Chen2020ComputationalEnvironments} differently from random disorder. Yet once the chains are localised enough, momentum rejuvenation's influence is negligible and the optimal dephasing rate is determined only by the IPR, as can be seen for $\eta = 1, 10$. Therefore gradient-induced localisation and disorder-induced localisation only have different effects on ENAQT when the gradients are strong enough to shorten the length scale over which momentum rejuvenation acts, but weak enough to ensure it is still present. 

\begin{figure}[H]
    \centering
    \includegraphics[width = .95\linewidth]{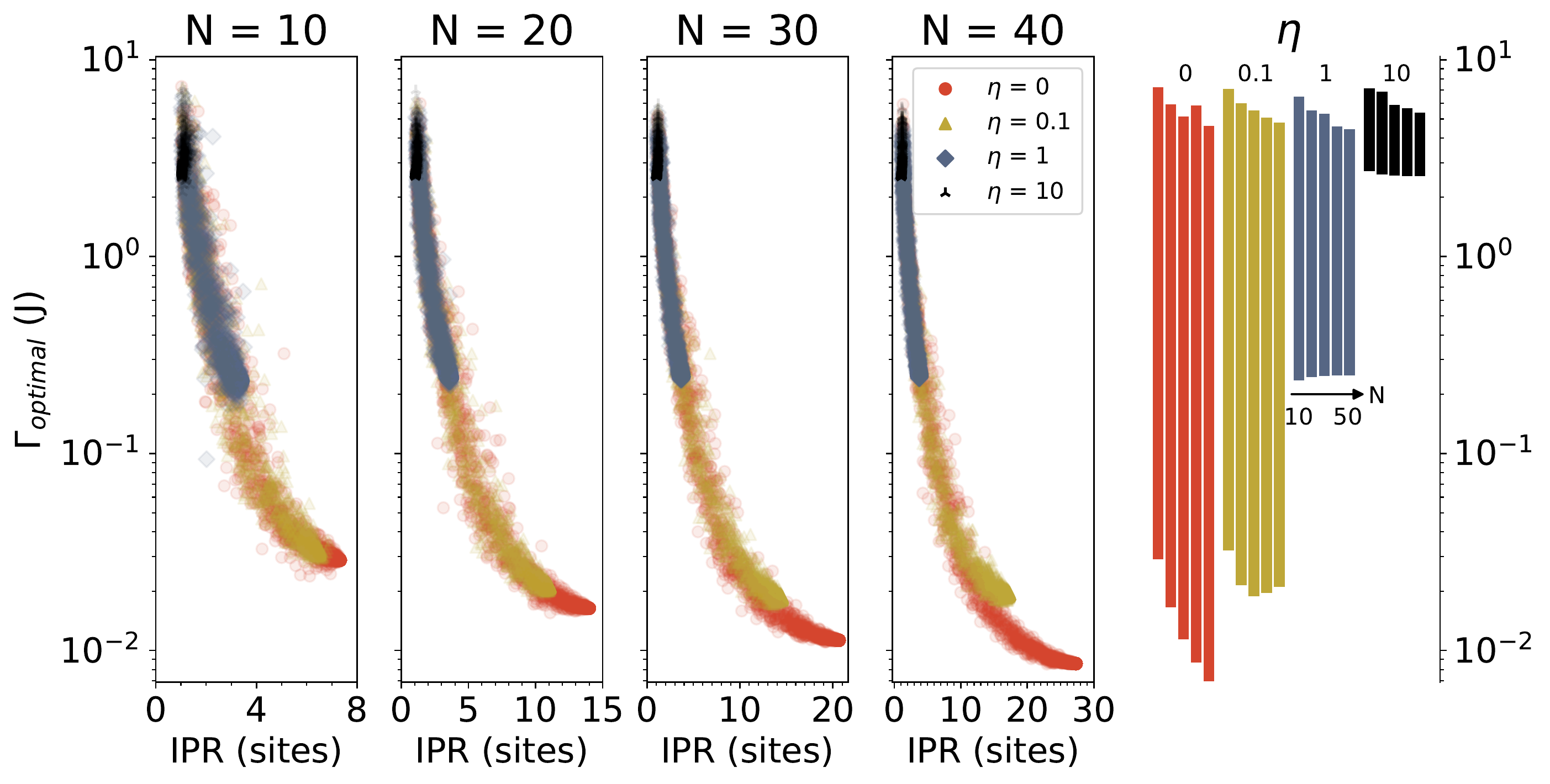}
    \caption{$\Gamma_{optimal}$ vs IPR for 4 different chain lengths, calculated with 100 realisations of disorder for each combination of gradients and disorder strength. As $N$ increases the same $\Gamma_{optimal}$-IPR response generally occurs, just stretched over a larger range of IPRs. The final panel shows how the range of $\Gamma_{optimal}$ for each gradient varies with length. For this final panel $N = 50$ was also considered, using 25 different realisations of disorder for each combination of disorder and energy gradient. The upper limit of each bar shows the optimal dephasing for the most localised chains, while the lower limit shows this for ordered chains where $\sigma = 0$. The momentum rejuvenation model predicts that as a system gets larger, the optimal dephasing rate decreases, we observe this for the lower edge of the $\eta = 0$ bars and partially for $\eta = 0.1$ before its effectiveness is reduced at larger $N$ as discussed in section \ref{sec: dephasing}. For the stronger gradients the lower range of $\Gamma_{optimal}$ does not significantly change with $N$, confirming that finite size effects are effectively suppressed for these chains, leaving only the effects of site to site detunings.}
    \label{fig: N10-40 gradient enaqt curves}
\end{figure}

The relative impact of momentum rejuvenation is then further affected by the size of the system itself. As discussed above, gradients reduce the maximum length over which momentum rejuvenation acts. This leads to differences in $\Gamma_{optimal}$ if that length is less than the system length. By extension we should then expect the differences between $\eta = 0$ and $\eta = 0.1$ to scale with the number of chain sites $N$. We can observe this directly in figure \ref{fig: N10-40 gradient enaqt curves} which demonstrates how the localising effects of energy gradients and random disorder affect ENAQT for chains of different lengths.

The upper limit of the $\Gamma_{optimal}$ bars in figure \ref{fig: N10-40 gradient enaqt curves}'s final panel shows consistent scaling behaviour across length, that is independent of gradients. This can be partially explained by regression to the mean as disorders are sampled from a Gaussian distribution. Maximally localised systems have large detunings between all sites, so the longer your system, the more detunings there are to maximise. Therefore, the larger the system, the harder it is to localise. Close inspection confirms this: the minimum IPR increases with chain length, and by extension the highest $\Gamma_{optimal}$ decreases with chain length.

We now focus our attention on the lower end of these $\Gamma_{optimal}$ ranges. First we note that the high gradient behaviour ($\eta = 1, 10$) has consistent lower limits for all chain lengths considered, meaning ENAQT is only determined by average site to site detunings, with little if any sensitivity to size. Momentum rejuvenation suggests that the larger the system, the lower its optimal noise rate, and we observe the zero gradient data extends to lower and lower dephasing rates as $N$ increases, exactly as predicted~\cite{Li2015MomentumFlow}. We note that the difference in $\Gamma_{optimal}$ between $\eta = 0$ and $\eta = 0.1$ increases with $N$, indicating that the range of which momentum rejuvenation acts has a finite length at $\eta = 0.1$ and so becomes less effective as system size increases. This can also be seen in how $\Gamma_{optimal}$ changes with $N$. The lower range of $\Gamma_{optimal}$, where $\eta = 0.1$ initially decreases as expected with momentum rejuvenation, then the trend reverses as increasing $N$ reduces the impact of momentum rejuvenation.

So far we have described under what conditions the size dependent effects of momentum rejuvenation can be observed, given the presence of energy gradients and random disorder. We now focus on the $\eta = 0$ limit we can directly capture how static disorder alters the influence of finite size effects on ENAQT. We consider the $N = 10-40$ chains and fit the $\eta = 0$ data with a power law of the form
\begin{equation}
\Gamma_{optimal}(\textnormal{IPR}) \propto \text{IPR}^{\lambda + \kappa \cdot \textnormal{IPR}}. 
\label{eqn: curve-power-law}
\end{equation}
The exponent $\lambda$ captures the response across all IPR values, corresponding to the influence of the invariant subspace and the need for line broadening. Meanwhile, the exponent $\kappa$ captures a varying influence, being negligible for very localised systems and most influential for systems with large IPR, capturing the influence of finite size effects such as momentum rejuvenation. We note that equation \ref{eqn: curve-power-law} is simply a phenomenological fit that best captures the data produced by our results, the data is not well fit by a single exponential and alternative functional forms likely require additional fitting parameters. As we show in table \ref{tab:fitting parameters} both $\lambda$ and $\kappa$ scale monotonically with chain length as expected. Further details and plots are presented in \ref{sec: curve fitting}.

\begin{table}[H]
    \centering
    \begin{tabular}{|c||c|c|c|c| }
    \hline
        N & $A$ $(J)$ & $\lambda$ & $\kappa$ & SD $(\times 10^{-3})$  \\ \hline
        10 & 1.59 & -3.14 & 0.07 & 1.80\\%\hline
        20 & 1.70 & -2.69 & 0.03 & 0.55\\%\hline
        30 & 1.74 & -2.51 & 0.02 & 0.33\\%\hline
        40 & 1.73 & -2.36 & 0.01 & 0.22\\\hline
    \end{tabular}
    \caption{Table of best fit values and standard deviation for each chain length with $\eta = 0$. $A$ captures the proportionality, $\lambda$ the size independent response, and $\kappa$ the size-dependent response. As chains get longer, the fits gets more accurate, and the parameters change monotonically as the same behaviours stretch over a new range of IPRs.}
    \label{tab:fitting parameters}
\end{table}

\subsection{Finite temperature Bloch Redfield model}
\label{sec: bloch-redfield}
\begin{figure}[H]
    \centering
    \includegraphics[width = .95\linewidth]{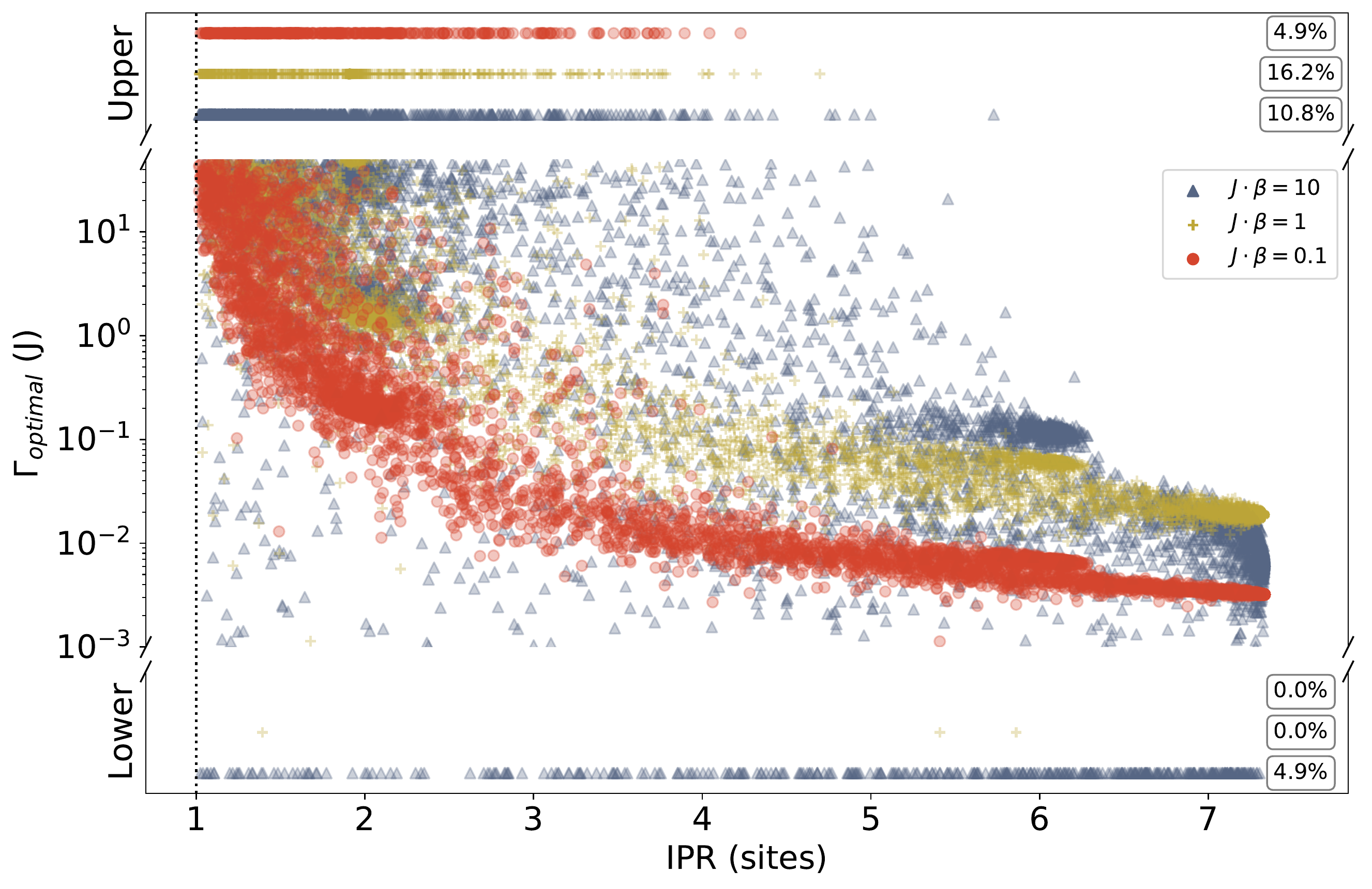}
    \caption{$\Gamma_{optimal}$ vs IPR for $N = 10$, considered at 3 inverse temperatures, once again for 100 realisations of disorder at each combination of gradient and disorder strengths. High and intermediate temperatures have broadly the same monotonic form, we note that the cooler the system is, the greater the $\Gamma_{optimal}$. These peaks are found using a bounded peak finding function, with the range $0 < \Gamma < 50J$. In the finite temperature limit we find some data points clustered at the edges of this range, suggesting either monotonic $I_{ss}$ vs $\Gamma$ curves, or $\Gamma_{optimal} \geq 50J$. We cut off all results within $10^{-3}J$ of either limit and collect them in the offset sections above and below the central axes. Each offset series is separated into points corresponding to different temperatures and annotated with a percentage to illustrate what fraction of data points corresponding to each temperature lie there.}
    \label{fig: 3 temps N10}
\end{figure}

As described in section \ref{sec: dynamics and rates}, we can go from  phenomenological pure dephasing model---effectively an infinite temperature limit---to a  microscopically-founded  finite-temperature approach with the use of the full, nonsecular Bloch-Redfield master equation, equation \ref{eqn: redfield ME}. These calculations are done against a flat spectral density for direct comparison with the pure dephasing results above. We define the inverse temperature $\beta = \frac{1}{k_B T}$, and consider 3 temperatures, $J \cdot \beta = 10, 1, 0.1$ (low, medium and high respectively), giving the results in figure \ref{fig: 3 temps N10}\footnotemark. 

\footnotetext{In $<$ 0.1\% of cases we found the steady state solver would fail, the optimisation procedure handled this by moving to the next trial point and continuing.}
Under these conditions we still recover the characteristic trend of a monotonic relationship between $\Gamma_{optimal}$ and IPR, and we report similar results for sufficiently wide non-flat spectra in figure \ref{fig: finite-temp-lorentzians}. We note that as temperatures lower, $\Gamma_{optimal}$ for a given IPR increases. As temperatures decrease, the specific energy landscape of each chain becomes more important ~\cite{Davidson2021PrinciplesAnalysis}, as it becomes harder to avoid trapping population in energy minima. As a result, the range of $\Gamma_{optimal}$ associated with any IPR gets broader continuously as temperatures get lower.% (for more detail on finite temperature effects see \ref{sec: finite temperature breakdown of curve}). 

We therefore conclude that the general ENAQT response to disorder depends not just on localisation, but also on avoiding the trapping of population in energetic minima. So when transfer rates up and down in energy become significantly different the chain population cannot explore all the system sites, trapping population in energetic minima. In this limit the universal response observed for pure dephasing breaks down, producing a regime which is very sensitive to the specifics of the energy landscape. By the reverse argument, if energy can move reasonably around a system then the monotonic relationship between optimal environmental coupling and IPR is well-defined.

\section{Conclusion}
\label{sec: conclusion}
We have systematically shown how localisation and optimal ENAQT are related for 1D chains, producing a universal trend strongly determined by the IPR. The IPR in turn is determined by an interplay of energy gradients, random disorder and system length. Comparing the range of $\Gamma_{optimal}$ for various lengths of chain provided further insight into how strong gradients can suppress the influence of finite size effects. Additionally we have found that steady state current in unbiased, disorder systems can be described by a power law containing size dependent and size independent contributions, illustrating that finite size effects such as momentum rejuvenation still affect how ENAQT acts on disordered systems.

Extending the model to include finite temperatures shows that the same response holds for high to intermediate temperatures. By contrast, at lower temperatures population can become trapped in local energetic minima. This decouples transport efficiency from eigenstate localisation and the transport becomes more sensitive to a chain's specific energy landscape. 

Throughout this paper we have shown that the localisation of a system's eigenstates is directly connected to what its optimal conditions for ENAQT are. By considering this for a large range of possible conditions we have developed new and broadly applicable insights into how localisation and finite size effects alter ENAQT in 1D. More work is required to confirm if this response is altered for higher dimensional systems where eigenstates may be further delocalised. For example simple tight-binding honeycomb lattices as found in graphene nano-ribbons can display quantum chaotic properties under weak static fields~\cite{Kolovsky2013Wannier-StarkLattice}, opening up a new class of system. The effects of localisation could be further investigated, whether taking a more fine-grained look at the unusual Wannier-Stark behaviour in \ref{sec: gradient vs ipr}, or going to much larger system sizes in order to limit the influence of finite size effects. Lastly, quasiperiodic systems such as the Aubry-André model could be considered, where transient effects such as stochastic resonance with anti-localised eigenstates~\cite{Gholami2017NoiseModels} may provide new insights into ENAQT beyond the steady state.

\section*{Acknowledgements}
We thank Scott Davidson, Dominic Rouse and Gerard Valentí-Rojas for the helpful discussions. This work was supported by EPSRC Grant No. EP/L015110/1. Computations were carried out using QuTiP~\cite{Johansson2013QuTiPSystems}, figures made in matplotlib~\cite{Hunter:2007}.

\bibliographystyle{unsrt}  
\bibliography{references}  %%% Remove comment to use the external .bib file (using bibtex).

\begin{thebibliography}{10}

\bibitem{Kassal2013DoesPhotosynthesis}
Ivan Kassal, Joel Yuen-Zhou, and Saleh Rahimi-Keshari.
\newblock {Does coherence enhance transport in photosynthesis?}
\newblock {\em Journal of Physical Chemistry Letters}, 4(3):362--367, 2013.

\bibitem{Ishizaki2009UnifiedApproach}
Akihito Ishizaki and Graham~R. Fleming.
\newblock {Unified treatment of quantum coherent and incoherent hopping
  dynamics in electronic energy transfer: Reduced hierarchy equation approach}.
\newblock {\em Journal of Chemical Physics}, 130(23), 2009.

\bibitem{Engel2007EvidenceSystems}
Gregory~S. Engel, Tessa~R. Calhoun, Elizabeth~L. Read, Tae-Kyu~Kyu Ahn, Tomáš
  Man{\v{c}}al, Yuan-Chung~Chung Cheng, Robert~E. Blankenship, and Graham~R.
  Fleming.
\newblock {Evidence for wavelike energy transfer through quantum coherence in
  photosynthetic systems}.
\newblock {\em Nature}, 446(7137):782--786, 4 2007.

\bibitem{Brixner2017ExcitonSystems}
Tobias Brixner, Richard Hildner, Jürgen K{\"{o}}hler, Christoph Lambert, and
  Frank W{\"{u}}rthner.
\newblock {Exciton Transport in Molecular Aggregates – From Natural Antennas
  to Synthetic Chromophore Systems}.
\newblock {\em Advanced Energy Materials}, 7(16):1--33, 2017.

\bibitem{Kundu2017NanoscaleHarvesting}
Simanta Kundu and Amitava Patra.
\newblock {Nanoscale strategies for light harvesting}.
\newblock {\em Chemical Reviews}, 117(2):712--757, 1 2017.

\bibitem{Amarnath2016MultiscalePlants}
Kapil Amarnath, Doran I.G.~G. Bennett, Anna~R. Schneider, and Graham~R.
  Fleming.
\newblock {Multiscale model of light harvesting by photosystem II in plants}.
\newblock {\em Proceedings of the National Academy of Sciences},
  113(5):1156--1161, 2016.

\bibitem{Bennett2013ADescription}
Doran I.~G. Bennett, Kapil Amarnath, and Graham~R. Fleming.
\newblock {A structure-based model of energy transfer reveals the principles of
  light harvesting in Photosystem II supercomplexes Methods : Extended
  Description}.
\newblock {\em Journal of the American Chemical Society}, 135(1):1--16, 6 2013.

\bibitem{Plenio2008Dephasing-assistedBiomolecules}
M.~B. Plenio and S.~F. Huelga.
\newblock {Dephasing-assisted transport: Quantum networks and biomolecules}.
\newblock {\em New Journal of Physics}, 10(11):113019, 11 2008.

\bibitem{Mohseni2008Environment-AssistedTransfer}
Masoud Mohseni, Patrick Rebentrost, Seth Lloyd, and Alán Aspuru-Guzik.
\newblock {Environment-assisted quantum walks in photosynthetic energy
  transfer}.
\newblock {\em Journal of Chemical Physics}, 129(17), 2008.

\bibitem{Chin2010Noise-assistedComplexes}
A.~W. Chin, A.~Datta, F.~Caruso, S.~F. Huelga, and M.~B. Plenio.
\newblock {Noise-assisted energy transfer in quantum networks and
  light-harvesting complexes}.
\newblock {\em New Journal of Physics}, 12(6):065002, 6 2010.

\bibitem{Zerah-Harush2018UniversalNetworks}
Elinor Zerah-Harush and Yonatan Dubi.
\newblock {Universal Origin for Environment-Assisted Quantum Transport in
  Exciton Transfer Networks}.
\newblock {\em Journal of Physical Chemistry Letters}, 9(7):1689--1695, 4 2018.

\bibitem{Dwiputra2021Environment-assistedEdges}
Donny Dwiputra and Freddy~P. Zen.
\newblock {Environment-assisted quantum transport and mobility edges}.
\newblock {\em Physical Review A}, 104(2):022205, 8 2021.

\bibitem{Lambert2012QuantumBiology}
Neill Lambert, Yueh-Nan Chen, Yuan-Chung Cheng, Che-Ming Li, Guang-Yin Chen,
  and Franco Nori.
\newblock {Quantum biology}.
\newblock {\em Nature Physics}, 9, 2012.

\bibitem{Huelga2013VibrationsBiology}
S.F. Huelga and M.B. Plenio.
\newblock {Vibrations, quanta and biology}.
\newblock {\em Contemporary Physics}, 54(4):181--207, 7 2013.

\bibitem{Stones2016VibronicTransfer}
Richard Stones and Alexandra Olaya-castro.
\newblock {Vibronic Coupling as a Design Principle to Optimize Photosynthetic
  Energy Transfer}.
\newblock {\em CHEMPR}, 1(6):822--824, 2016.

\bibitem{Harush2021DoNot}
Elinor~Zerah Harush and Yonatan Dubi.
\newblock {Do photosynthetic complexes use quantum coherence to increase their
  efficiency? Probably not}.
\newblock {\em Science Advances}, 7(8):1--9, 2021.

\bibitem{Higgins2021PhotosynthesisTransfer}
Jacob~S Higgins, Lawson~T Lloyd, Sara~H Sohail, Marco~A Allodi, John~P Otto,
  Rafael~G Saer, Ryan~E Wood, Sara~C Massey, Po~Chieh Ting, Robert~E
  Blankenship, and Gregory~S Engel.
\newblock {Photosynthesis tunes quantum-mechanical mixing of electronic and
  vibrational states to steer exciton energy transfer}.
\newblock {\em Proceedings of the National Academy of Sciences of the United
  States of America}, 118(11), 3 2021.

\bibitem{Duan2017NatureTransfer.}
Hong-Guang Duan, Valentyn~I Prokhorenko, Richard~J Cogdell, Khuram Ashraf,
  Amy~L Stevens, Michael Thorwart, and R~J~Dwayne Miller.
\newblock {Nature does not rely on long-lived electronic quantum coherence for
  photosynthetic energy transfer.}
\newblock {\em Proceedings of the National Academy of Sciences of the United
  States of America}, 114(32):8493--8498, 8 2017.

\bibitem{Li2015MomentumFlow}
Ying Li, Filippo Caruso, Erik Gauger, and Simon~C. Benjamin.
\newblock {Momentum rejuvenation underlies the phenomenon of noise-assisted
  quantum energy flow}.
\newblock {\em New Journal of Physics}, 17(1):13057, 2015.

\bibitem{Zerah-Harush2020EffectsTransport}
Elinor Zerah-Harush and Yonatan Dubi.
\newblock {Effects of disorder and interactions in environment assisted quantum
  transport}.
\newblock {\em Physical Review Research}, 2(2):23294, 2020.

\bibitem{Anderson1958AbsenceLattices}
P.~W. Anderson.
\newblock {Absence of diffusion in certain random lattices}.
\newblock {\em Physical Review}, 109(5):1492--1505, 1958.

\bibitem{Wannier1960WaveField}
Gregory~H. Wannier.
\newblock {Wave functions and effective hamiltonian for bloch electrons in an
  electric field}.
\newblock {\em Physical Review}, 117(2):432--439, 1960.

\bibitem{Caruso2009HighlyTransport}
F.~Caruso, A.~W. Chin, A.~Datta, S.~F. Huelga, and M.~B. Plenio.
\newblock {Highly efficient energy excitation transfer in light-harvesting
  complexes: The fundamental role of noise-assisted transport}.
\newblock {\em Journal of Chemical Physics}, 131(10), 2009.

\bibitem{Spano1991CooperativeAggregates}
Francis~C. Spano, Jan~R. Kuklinski, and Shaul Mukamel.
\newblock {Cooperative radiative dynamics in molecular aggregates}.
\newblock {\em The Journal of Chemical Physics}, 94(11):7534--7544, 1991.

\bibitem{Gulli2019MacroscopicNanotubes}
Marco Gull{\`{i}}, Alessia Valzelli, Francesco Mattiotti, Mattia Angeli, Fausto
  Borgonovi, and Giuseppe~Luca Celardo.
\newblock {Macroscopic coherence as an emergent property in molecular
  nanotubes}.
\newblock {\em New Journal of Physics}, 21(1), 2019.

\bibitem{Strumpfer2012HowLight-harvesting}
J.~Str{\"{u}}mpfer, M.~{\c{S}}ener, and K.~Schulten.
\newblock {How quantum coherence assists photosynthetic light-harvesting}.
\newblock {\em Journal of Physical Chemistry Letters}, 3(4):536--542, 2012.

\bibitem{Jurcevic2014QuasiparticleSystem}
P.~Jurcevic, B.~P. Lanyon, P.~Hauke, C.~Hempel, P.~Zoller, R.~Blatt, and C.~F.
  Roos.
\newblock {Quasiparticle engineering and entanglement propagation in a quantum
  many-body system}.
\newblock {\em Nature}, 511(7508):202--205, 2014.

\bibitem{Levitov1989AbsenceInteraction}
L.~S. Levitov.
\newblock {Absence of Localization of Vibrational Modes Due to Dipole-Dipole
  Interaction}.
\newblock {\em European Physics Letters}, 9(1):83--86, 1989.

\bibitem{Evers2008AndersonTransitions}
Ferdinand Evers and Alexander~D. Mirlin.
\newblock {Anderson transitions}.
\newblock {\em Reviews of Modern Physics}, 80(4):1355--1417, 2008.

\bibitem{Celardo2016ShieldingHopping}
G.~L. Celardo, R.~Kaiser, and F.~Borgonovi.
\newblock {Shielding and localization in the presence of long-range hopping}.
\newblock {\em Physical Review B}, 94(14):1--12, 2016.

\bibitem{Chavez2020Disorder-EnhancedCavities}
Nahum~C. Ch{\'{a}}vez, Francesco Mattiotti, J.~A. M{\'{e}}ndez-Berm{\'{u}}dez,
  Fausto Borgonovi, and G.~Luca~Celardo.
\newblock {Disorder-Enhanced and Disorder-Independent Transport with long-range
  hopping: application to molecular chains in optical cavities}.
\newblock {\em arXiv}, pages 1--25, 2020.

\bibitem{Gholami2017NoiseModels}
Ehsan Gholami and Zahra~Mohammaddoust Lashkami.
\newblock {Noise, delocalization, and quantum diffusion in one-dimensional
  tight-binding models}.
\newblock {\em Physical Review E}, 95(2):1--10, 2017.

\bibitem{Lorenzo2018RemnantsNoise}
S.~Lorenzo, T.~Apollaro, G.~M. Palma, R.~Nandkishore, A.~Silva, and J.~Marino.
\newblock {Remnants of Anderson localization in prethermalization induced by
  white noise}.
\newblock {\em Physical Review B}, 98(5):1--6, 2018.

\bibitem{Zhu2021ProbingComputer}
D.~Zhu, S.~Johri, N.~H. Nguyen, C.~Huerta Alderete, K.~A. Landsman, N.~M.
  Linke, C.~Monroe, and A.~Y. Matsuura.
\newblock {Probing many-body localization on a noisy quantum computer}.
\newblock {\em Physical Review A}, 103(3):1--7, 2021.

\bibitem{Malla2018SpinfulCoupling}
Rajesh~K Malla and M~E Raikh.
\newblock {Spinful Aubry-Andr{\'{e}} model in a magnetic field: Delocalization
  facilitated by a weak spin-orbit coupling}.
\newblock {\em Physical Review B}, 97(21):1--11, 2018.

\bibitem{Bonca2018DynamicsBaths}
Janez Bon{\v{c}}a, Stuart~A. Trugman, and Marcin Mierzejewski.
\newblock {Dynamics of the one-dimensional Anderson insulator coupled to
  various bosonic baths}.
\newblock {\em Physical Review B}, 97(17):1--8, 2018.

\bibitem{Prelovsek2018TransientBosons}
P.~Prelov{\v{s}}ek, J.~Bon{\v{c}}a, and M.~Mierzejewski.
\newblock {Transient and persistent particle subdiffusion in a disordered chain
  coupled to bosons}.
\newblock {\em Physical Review B}, 98(12):1--6, 2018.

\bibitem{Deutsch2018EigenstateHypothesis}
Joshua~M. Deutsch.
\newblock {Eigenstate thermalization hypothesis}.
\newblock {\em Reports on Progress in Physics}, 81(8), 5 2018.

\bibitem{DAlessio2016FromThermodynamics}
Luca D'Alessio, Yariv Kafri, Anatoli Polkovnikov, and Marcos Rigol.
\newblock {From quantum chaos and eigenstate thermalization to statistical
  mechanics and thermodynamics}.
\newblock {\em Advances in Physics}, 65(3):239--362, 2016.

\bibitem{Sa2020ComplexChaos}
Lucas S{\'{a}}, Pedro Ribeiro, and TomaŽ Prosen.
\newblock {Complex Spacing Ratios: A Signature of Dissipative Quantum Chaos}.
\newblock {\em Physical Review X}, 10(2):1--23, 2020.

\bibitem{Rubio-Garcia2021FromLiouvillians}
Álvaro Rubio-Garc{\'{i}}a, Rafael~A. Molina, and Jorge Dukelsky.
\newblock {From integrability to chaos in quantum Liouvillians}.
\newblock {\em arXiv}, 2021.

\bibitem{Kleinman1990CommentLocalization}
Leonard Kleinman.
\newblock {Comment on existence of Wannier-Stark localization}.
\newblock {\em Physical Review B}, 41(6):3857--3858, 1990.

\bibitem{Emin1987Phonon-assistedField}
David Emin and C.~F. Hart.
\newblock {Phonon-assisted hopping of an electron on a Wannier-Stark ladder in
  a strong electric field}.
\newblock {\em Physical Review B}, 36(5):2530--2546, 1987.

\bibitem{Wilkinson1996ObservationPotential}
S.~R. Wilkinson, C.~F. Bharucha, K.~W. Madison, Qian Niu, and M.~G. Raizen.
\newblock {Observation of atomic Wannier-Stark ladders in an accelerating
  optical potential}.
\newblock {\em Physical Review Letters}, 76(24):4512--4515, 1996.

\bibitem{Johansson2013QuTiPSystems}
J.~R. Johansson, P.~D. Nation, and Franco Nori.
\newblock {QuTiP 2: A Python framework for the dynamics of open quantum
  systems}.
\newblock {\em Computer Physics Communications}, 2013.

\bibitem{Jeske2015Bloch-RedfieldComplexes}
Jan Jeske, David~J. Ing, Martin~B. Plenio, Susana~F. Huelga, and Jared~H. Cole.
\newblock {Bloch-Redfield equations for modeling light-harvesting complexes}.
\newblock {\em Journal of Chemical Physics}, 142(6), 2015.

\bibitem{Huo2012InfluenceSystems}
Pengfei Huo and David~F. Coker.
\newblock {Influence of environment induced correlated fluctuations in
  electronic coupling on coherent excitation energy transfer dynamics in model
  photosynthetic systems}.
\newblock {\em Journal of Chemical Physics}, 136(11), 2012.

\bibitem{Eastham2016Bath-inducedApproximation}
P.~R. Eastham, P.~Kirton, H.~M. Cammack, B.~W. Lovett, and J.~Keeling.
\newblock {Bath-induced coherence and the secular approximation}.
\newblock {\em Physical Review A}, 94(1):1--9, 2016.

\bibitem{Breuer2002TheSystems}
Heinz~Peter Breuer and Francesco Petruccione.
\newblock {\em {The Theory of Open Quantum Systems}}.
\newblock Oxford University Press, 2002.

\bibitem{Brumer2018SheddingProcesses}
Paul Brumer.
\newblock {Shedding (Incoherent) Light on Quantum Effects in Light-Induced
  Biological Processes}.
\newblock {\em Journal of Physical Chemistry Letters}, 9(11):2946--2955, 2018.

\bibitem{Axelrod2018AnLight}
Simon Axelrod and Paul Brumer.
\newblock {An efficient approach to the quantum dynamics and rates of processes
  induced by natural incoherent light}.
\newblock {\em Journal of Chemical Physics}, 149(11), 2018.

\bibitem{Chaudhry2016AEffects}
Adam~Zaman Chaudhry.
\newblock {A general framework for the Quantum Zeno and anti-Zeno effects}.
\newblock {\em Scientific Reports}, 6(June):1--10, 2016.

\bibitem{Chen2020ComputationalEnvironments}
Ning Chen, Murali Devi, and Seogjoo~J. Jang.
\newblock {Computational modeling of charge hopping dynamics along a disordered
  one-dimensional wire with energy gradients in quantum environments}.
\newblock {\em The Journal of Chemical Physics}, 153(5):054109, 2020.

\bibitem{Davidson2021PrinciplesAnalysis}
Scott Davidson, Felix~A. Pollock, and Erik Gauger.
\newblock {Principles underlying efficient exciton transport unveiled by
  information-geometric analysis}.
\newblock {\em Physical Review Research}, 3(3):L032001, 7 2021.

\bibitem{Kolovsky2013Wannier-StarkLattice}
Andrey~R. Kolovsky and Evgeny~N. Bulgakov.
\newblock {Wannier-Stark states and Bloch oscillations in the honeycomb
  lattice}.
\newblock {\em Physical Review A}, 87(3):033602, 3 2013.

\bibitem{Hunter:2007}
John~D Hunter.
\newblock {Matplotlib: A 2D Graphics Environment}.
\newblock {\em Computing in Science {\&} Engineering}, 9(3):90--95, 2007.

\bibitem{Lhuillier2010QuantumTemperature}
E.~Lhuillier, I.~Ribet-Mohamed, A.~Nedelcu, V.~Berger, and E.~Rosencher.
\newblock {Quantum transport in weakly coupled superlattices at low
  temperature}.
\newblock {\em Physical Review B - Condensed Matter and Materials Physics},
  81(15):1--12, 2010.

\end{thebibliography}

\appendix

\section{Gradient-only localisation vs IPR}
\label{sec: gradient vs ipr}
\begin{figure}[H]
    \centering
    \includegraphics[width = .8\linewidth]{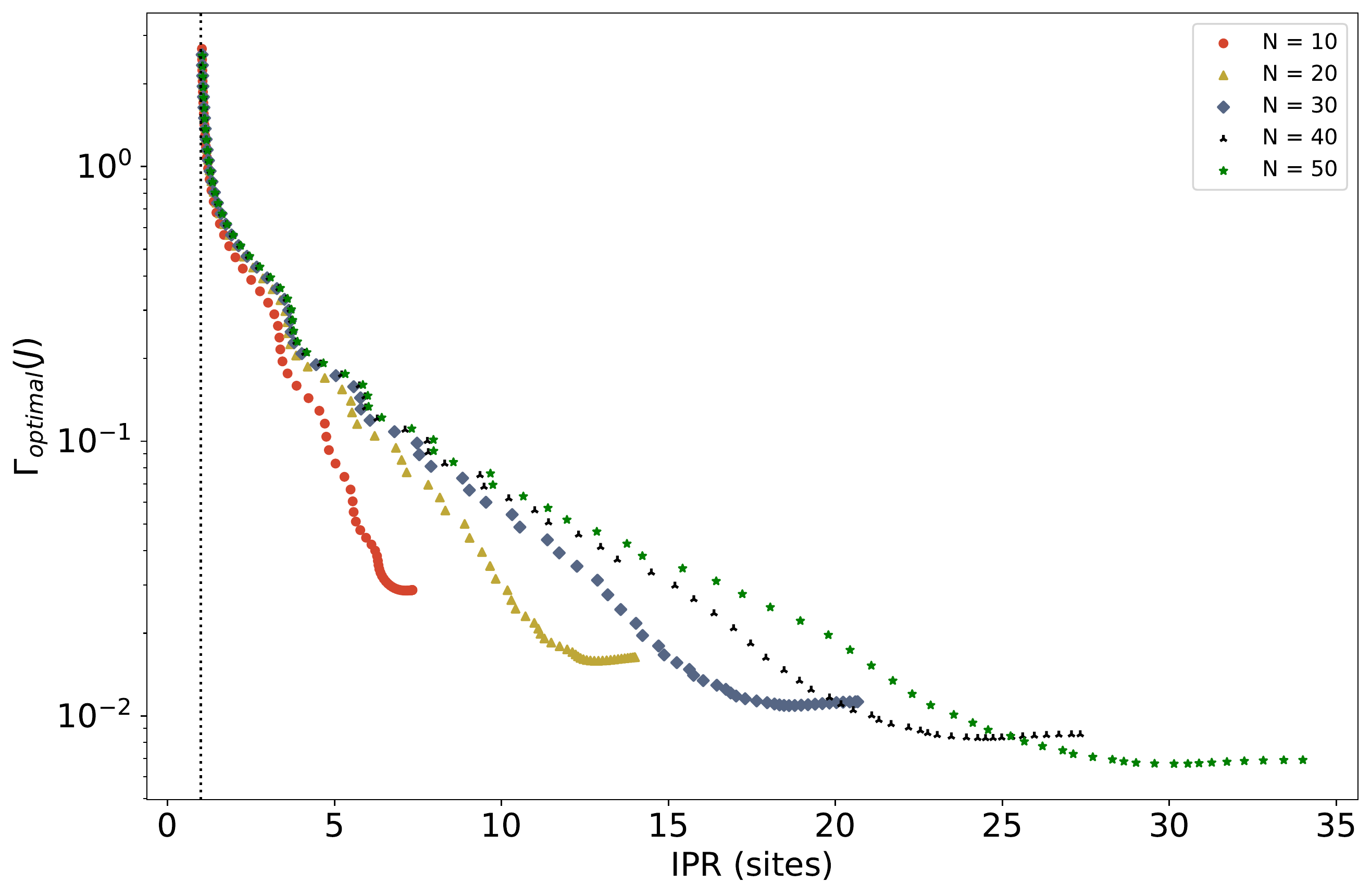}
    \caption{$\Gamma_{optimal}$ vs IPR when there is no applied disorder, only applied gradients. We see consistent stepped behaviour across all lengths as the Wannier-Stark localisation increases. }
    \label{fig: gradient vs ipr}
\end{figure}

In the limit where we apply a gradient but no disorder, we find a set of behaviours quite different from that described in the main body of the paper, we note these behaviours are only really visible on a log plot and for small enough rates that it would be very hard to observe them experimentally. Looking at figure \ref{fig: gradient vs ipr} we see a consistent and strangely stepped behaviour where $\Gamma_{optimal}$ increases irregularly as the IPR decreases. We speculate this may be due to the eigenstates under Wannier-Stark localisation being very consistent in spread and overlap. As a result of this ordering, changing the IPR implies not only further localising eigenstates but also consistently shifting the mutual participation any pair of eigenstates have on a common set of sites. This is in stark contrast to systems with random static disorder where there is no consistent mutual presence of eigenstates to disrupt.

Another potential explanation comes from prior experimental work on quantum transport in biased, weakly coupled superlattices~\cite{Lhuillier2010QuantumTemperature}. This work demonstrated the existence of a single plateau in the current-bias relationship, and found it could be physically explained by the influence of Wannier-Stark localisation acting against field-assisted tunneling between adjacent wells. More analysis is needed to determine if this could be connected to our work or if the resemblance is merely superficial. The use of more local measures for localisation, and studying larger systems would let one better distinguish the effect these gradients have on eigenstates in the bulk as compared to the edge.

\section{Curve fitting}
\label{sec: curve fitting}
\begin{figure}[H]
    \centering
    \includegraphics[width = .8\linewidth]{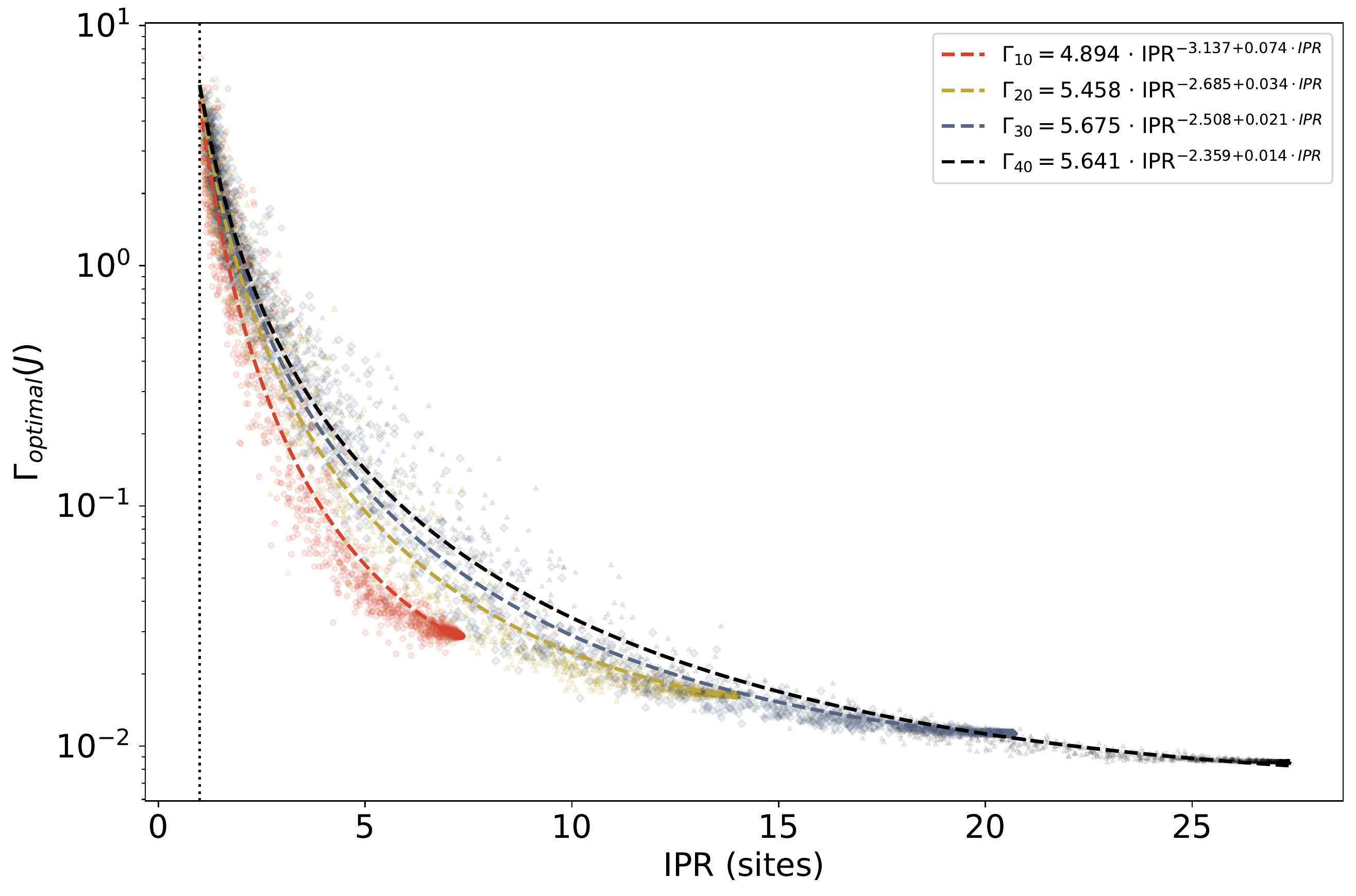}
    \caption{$\Gamma_{optimal}$ vs IPR with power laws fitting each dataset. To avoid the influence of gradients the data was only fit to subset of data with no field gradient applied. Parameters shown, and listed in table \ref{tab:fitting parameters}.}
    \label{fig: dephasing vs ipr fit}
\end{figure}

We have explained that the curves in figure~\ref{fig: N40 gradient comparison} show two different responses: a steep disordered one largely determined by exponential localisation~\cite{Anderson1958AbsenceLattices} and the invariant subspace, and a flatter response characterised by the greater presence of momentum rejuvenation and coherent transport. As we do not have an analytic form for the curves in figures \ref{fig: N40 gradient comparison} and \ref{fig: N10-40 gradient enaqt curves} we directly fit them, and find that the change in response across IPR is modelled well by a power law of the form 
\begin{equation}
    \Gamma_{optimal, N}(\textnormal{IPR})= A \cdot \text{IPR}^{\lambda + \kappa \cdot \textnormal{IPR}},
    \label{eqn: fit function}
\end{equation}
where the proportionality constant $A$ is determined by systematic factors such as length, coupling strength, injection and extraction rates etc, while $\lambda$ describes the constant response to localisation, and $\kappa$ captures the increasing presence of coherent effects as the IPR increases. As such $\lambda$ captures the ever present exponential localisation from disorder~\cite{Anderson1958AbsenceLattices}, $\kappa$ captures the varying presence of finite size effects and momentum rejuvenation. 

\begin{figure}[H]
    \centering
    \includegraphics[width = .75\linewidth]{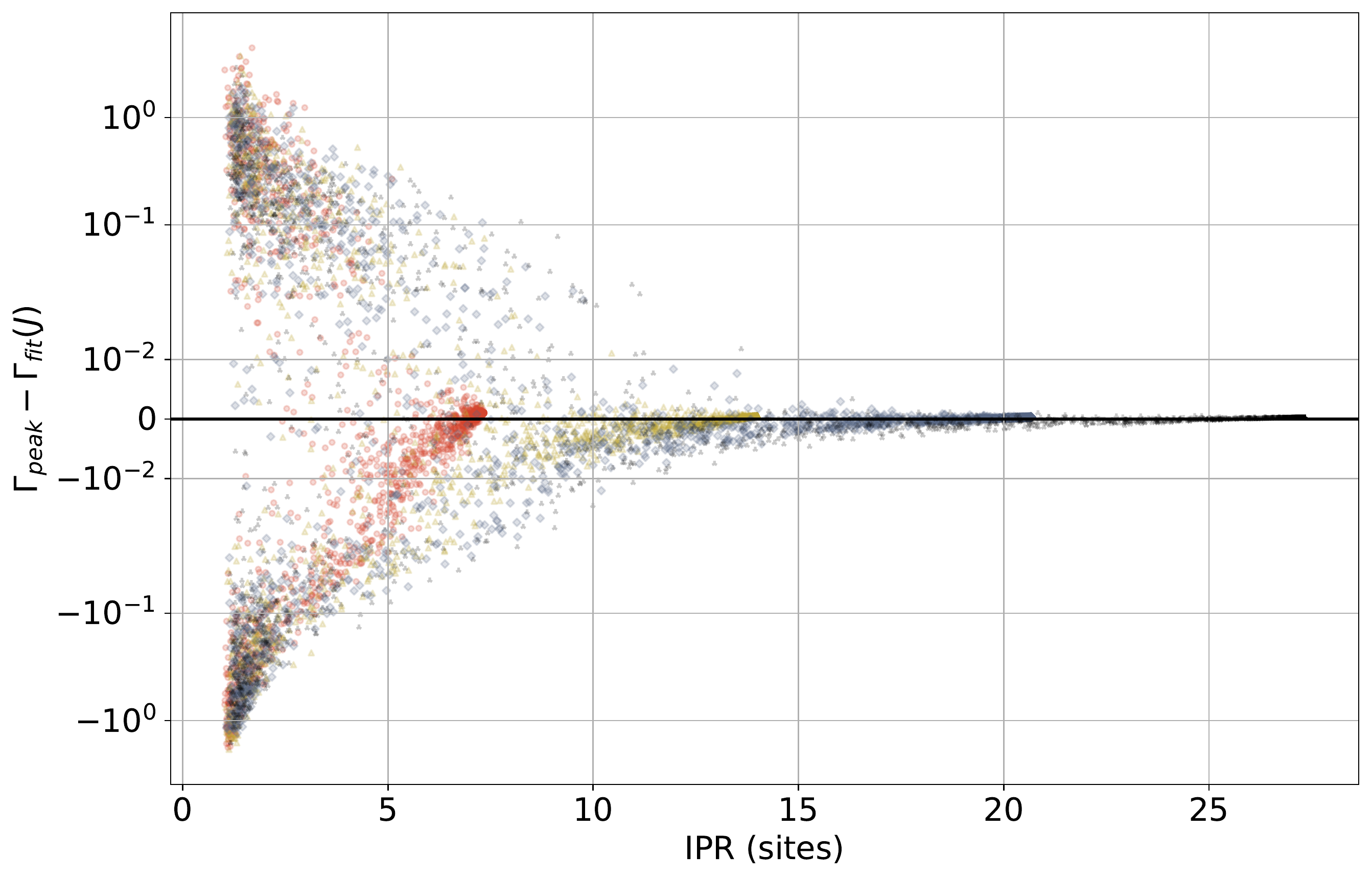}
    \caption{Residuals due to fitting power laws to data from various disorder strengths for N = 40.}
    \label{fig: fit residual}
\end{figure}

The best fits were found by using the logarithm of the data and equation \ref{eqn: fit function}. The best fit parameters are listed in table \ref{tab:fitting parameters}, and the fit residuals shown in figure \ref{fig: fit residual}. We find the fitting parameters scale monotonically with changes in $N$ as expected. The error is calculated using the quadrature sum of the covariance matrix diagonals generated by the fitting function. 

\section{Injecting on single sites}
\label{sec: single-site injection}
Throughout this work we have used a pumping scheme that injects population equally across all system sites. Here we show that response to localisation remains qualitatively the same when injecting onto a single site, by considering 2000 chains of length 20.
\begin{figure}[H]
    \centering
    \includegraphics[width = .75\linewidth]{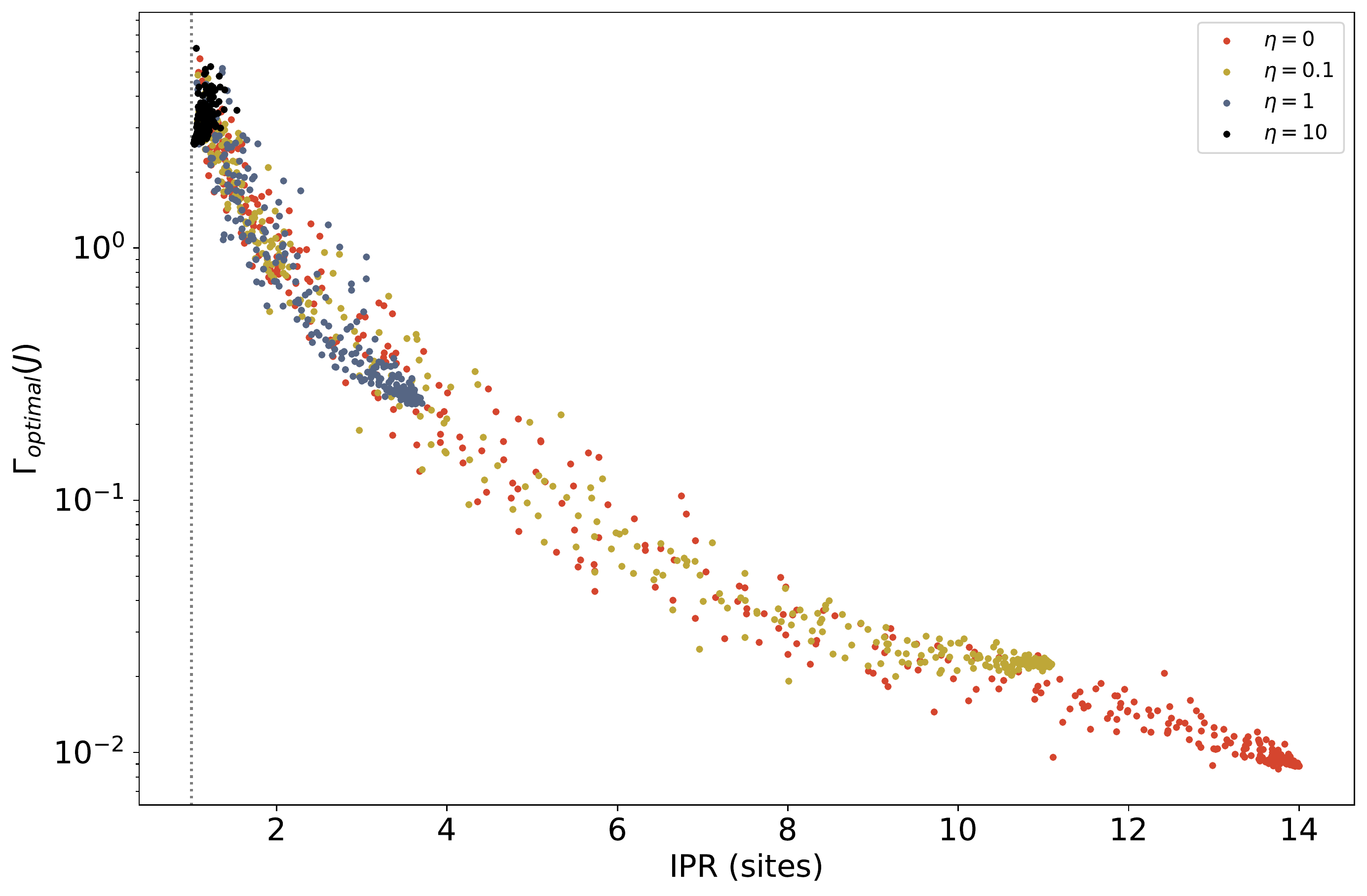}
    \caption{$\Gamma_{optimal}$ for 2000 length 20 chains under a range of Wannier-Stark and Anderson Localisation, the pumping scheme is extracting out the the $N^{th}$ site, and injecting on the $2^{nd}$ site. We see a very similar response to pumping on all sites, but expect the optimal dephasing rates to decrease the closer the injection site is to the extraction site.}
    \label{fig:single-inject20}
\end{figure}

\section{Population Uniformisation}
\label{sec: uniformisation}

There are multiple perspectives on ENAQT, and one prior, unified approach is provided by population uniformisation~\cite{Zerah-Harush2018UniversalNetworks}, which has also been studied in disordered systems~\cite{Zerah-Harush2020EffectsTransport}. The key measure used in this approach is the variance of steady state populations across all chain sites, uniformity is highest when variance is minimised. The population uniformisation theory predicts that this occurs at, or near to peak steady state currents. In other words $\Gamma_{min. var} \approx \Gamma_{optimal}$.

\begin{figure}[H]
    \centering
    \includegraphics[width = .8\linewidth]{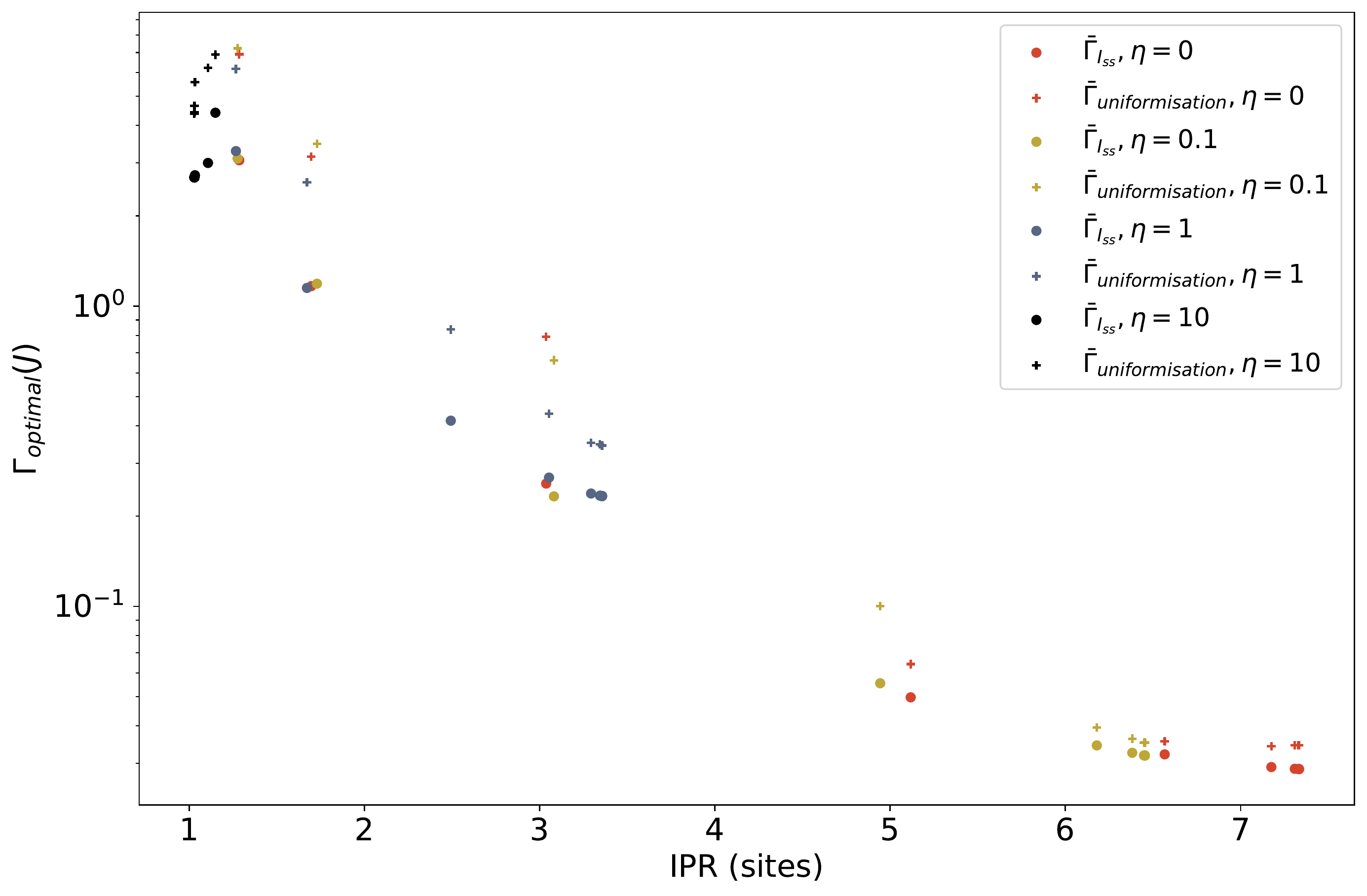}
    \caption{The averaged optimal dephasing rates with respect to the steady state current and population uniformisation respectively. Each point presented is the average of 50 realisations of disorder.}
    \label{fig: enaqt-uniformisation-comparison}
\end{figure}

We tested this by generating a representative sample of length 10 chains, then determined the optimal dephasing rates for steady state current and population uniformisation. After considering 50 realisations of disorder at each point in the parameter space and averaging, we indeed found that the two rates are correlated, agreeing very closely in ordered systems, and less so in highly disordered cases. As shown in figure \ref{fig: enaqt-uniformisation-comparison}, we find the average rates for peak uniformisation are consistently greater than those for the peak current. However, even in the divergent cases such as highly disordered systems, the on-site populations are still relatively uniform
when the steady state current is maximised. We also note that the difference between these two rates is mainly dependent on the IPR, and is insensitive to the gradient applied to the system. As such we can say that while offering a different view, our results are consistent with population uniformisation.

\section{Non-flat spectral densities}
\label{sec: drude-lorentz-spectra}

We mainly focus in this work on pure dephasing and non-peaked spectral densities to show the generic influence of localisation on optimal ENAQT conditions. Here we show the effects of a typical Drude-Lorentz spectral density \begin{equation}
    \mathcal{J}(\omega) = \lambda \cdot \frac{2}{\pi}\cdot \frac{\omega (\frac{1}{\tau})}{\omega^2 + (\frac{1}{\tau})^2},
\end{equation}
where $\frac{1}{\tau}$ is the Lorentzian linewidth and $\lambda$ is the coupling to the phonon bath.

\begin{figure}[H]
    \centering
    \includegraphics[width = .8\linewidth]{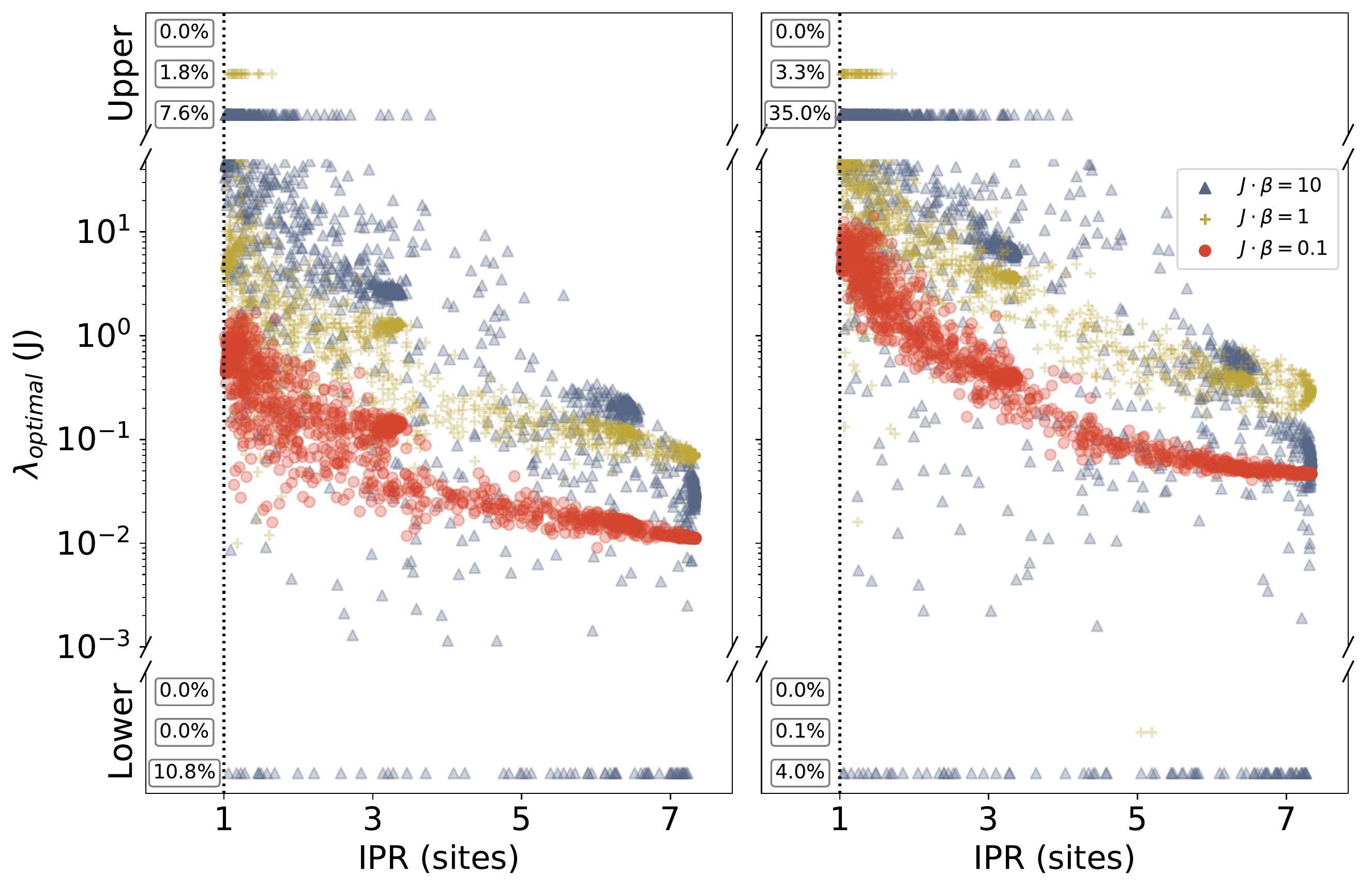}
    \caption{Optimal dephasing vs IPR with Drude-Lorentz spectral densities with linewidths $\tau^{-1} = 1J$ (left), and $10J$ (right). Note that as the linewidth increases by an order of magnitude, so do the peak phonon couplings $\lambda$. As a consequence, many more cases peak are above the upper limit of our peak finding code for the larger linewidth.}
    \label{fig: finite-temp-lorentzians}
\end{figure}

We consider multiple linewidths and their effects on 2000 chains, generated the same way as in the body of the text. We specifically use the linewidths $\tau^{-1} = 1, 10J$ to see the effect of linewidths equivalent to average system spacing, as shown in figure \ref{fig: finite-temp-lorentzians}. We do not consider much narrower linewidths as those would typically coincide with non-Markovian effects which are beyond the scope of the model we use.

We find that for these Lorentzians we still recover the same qualitative trends seen for flat spectral densities, though we note as $\tau^{-1}$ increases, the optimal phonon couplings increase as well. This is because as $\tau^{-1}$ increases, $\mathcal{J}(\omega)$ begins to scale with $\frac{\lambda}{\tau^{-1}}$. Thus for a system with an optimal set of transition rates, a larger linewidth means larger phonon couplings $\lambda$ are optimal.

\section{Transient effects}
\label{sec: transient effects}
Given recent interest and investigations into transient effects in tight-binding systems, we carried out dynamical calculations to check if effects beyond the steady state were contained in our model. In all cases we found that the dynamics converged onto the steady state behaviour, and remained converged to long time. Analysis of the chain Liouvillians confirmed that in all cases the steady state was uniquely defined, Hermitian and trace preserving. The same held true for the Bloch-Redfield results.

\begin{figure}[H]
    \centering
    \includegraphics[width=.75\linewidth]{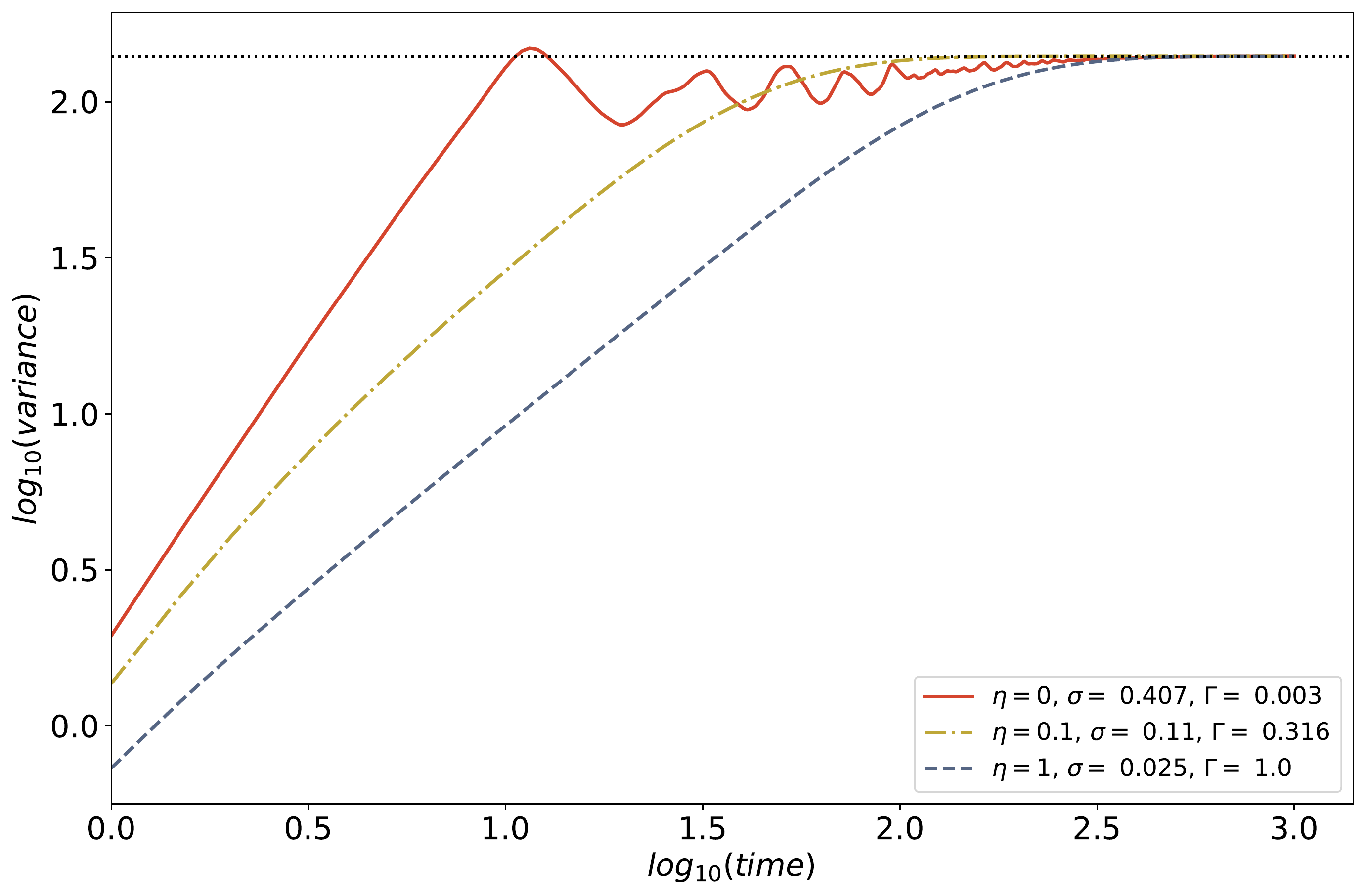}
    \caption{Variance of a localised initial wavefunction injected onto the centre of 3 representative length 41 chains with no injection or extraction. All eventually converge onto the steady state (horizontal dashed line). All quantities dimensionless.}
    \label{fig:closed-chain-variance-various}
\end{figure}

Taking inspiration from a prior study on stochastic resonance~\cite{Gholami2017NoiseModels} we investigated the variance of a time-evolving localised initial state injected onto the centre of a chain of 41 sites. This chain was closed, having no extraction or pumping, and the variance was defined as 
\begin{equation}
        variance = \sum_{n} n^2 |\psi_n|^2,
\end{equation}
where $n$ is the site index with respect to the central site, such that for an odd length system $n \in [-\frac{N-1}{2},\frac{N-1}{2}]$, $\psi_n$ is the normalised, initial state. We do not find any evidence of non-trivial transient effects in our model, instead as shown in figure \ref{fig:closed-chain-variance-various} we observe a set of smooth continuous approaches to the steady state, with some over- or under-damping depending on the dephasing rate. 
\end{document}